\documentstyle[aps,prl,epsfig]{revtex}

\newcommand{\be}{\begin{equation}}
\newcommand{\ee}{\end{equation}}
\newcommand{\bea}{\begin{eqnarray}}
\newcommand{\eea}{\end{eqnarray}}

\begin{document}
%\begin{titlepage}
%\vspace{1in}

\title
{\hfill{KEK-TH/809} \\
\bf Duality and Integrability of Two Dimensional String Effective Action}

%\vspace{1in}

%\normalsize

\author{ Ashok Das$^a$, J. Maharana$^{b}$\footnote{Permanent address:
Institute of Physics, Bhubaneswar 751005, India}  and  A. Melikyan$^a$}

\address{$^a$ Department of Physics and Astronomy,
University of Rochester, NY 14627-0171, USA}
\address{$^b$ Theory Division, KEK, 
Tsukuba, Ibaraki 3050801, Japan}

%\end{center}
\maketitle
\vskip .5cm

\begin{abstract}
We present a prescription for constructing the monodromy matrix,
$\widehat{\cal M}(\omega)$, for $O(d,d)$ invariant string effective
actions and derive its transformation properties under
the $T$-duality group. This allows us to construct $\widehat{\cal
M}(\omega)$ for new backgrounds, starting from known ones, which are related
by $T$-duality. As an application, we derive the monodromy matrix for
the exactly solvable Nappi-Witten model, both when $B=0$ and $B\neq 0$. 

\end{abstract}

\vspace{.7in}
 
%\end{titlepage}

%\section{INTRODUCTION }

%\setcounter{equation}{0}

%\def\theequation{\thesection.\arabic{equation}}

The symmetry structure of string theories is one of its most
fascinating features. In particular, duality symmetries have played a
very  important role in
our understanding of string dynamics \cite{gpr,as,jm}.
Whereas $T$-duality properties
can be subjected to tests within the perturbative frame work, the
$S$-duality symmetry explores  nonperturbative attributes of string
theories. The tree level string effective action, compactified on a
$d$-dimensional torus, $T^d$, is known to be invariant under the
noncompact global symmetry group $O(d,d)$. Indeed, this symmetry
property has been exploited, in the past, to generate new string vacua from 
known solutions.

The dimensionally reduced string effective action, in two space-time
dimensions, is known to possess enlarged symmetries, which has been explored
by several authors \cite{ib,jmi,jhs,sen}. 
The effective action is described by a nonlinear
$\sigma$-model, defined over a coset, which is coupled to gravity. 
In the past, two dimensional models, derived from dimensional
reduction of  higher dimensional Einstein gravity as well as
supergravity theories, have been studied to bring out their integrability
properties \cite{bm,nic1,nic2}.
In this
context, one may recall that the construction of the monodromy matrix
turns out to be one of the principal objectives in the study of integrable
systems \cite{nic3,das}. 
Furthermore, the integrability properties of such models have been
studied in the past in some detail from different perspectives.
 Recently, we have 
investigated the scattering of plane fronted waves, in four dimensions,
which  correspond to
massless states of closed strings \cite{dmm}. 
The scattering process is described
by an effective two dimensional action due to the presence of
isometries along the directions transverse to the axis of
collision. Therefore, the reduced action is endowed with an $O(2,2)$
symmetry, as is obvious from the preceding discussion. We were able to
derive the monodromy matrix for such a scattering process for the  specific
background configurations involved and had established
an intimate connection between the integrability properties of the model and
the underlying $O(2,2)$ symmetry.

The purpose of this note is to derive the monodromy matrix,
 $\widehat{\cal M}(\omega)$,  for a 2-dimensional
 string effective action, obtained  from
a $D$-dimensional effective action, which is compactified on $T^d$ and
thus enjoys isometries along $d=D-2$ spatial directions. We provide 
the prescription for the 
construction of the monodromy matrix, in this case, and  derive the
transformation property of $\widehat{\cal M}(\omega)$ under $O(d,d)$
transformations. In this process,  we are able to synthesize the classical
integrability properties and the duality symmetry of string theory in a
novel way. Furthermore, as an illustrative example, we 
construct the monodromy matrix for the well known Nappi-Witten model
 \cite{nw}, both for $B=0$ and $B\neq 0$, 
and demonstrate explicitly the transformation of $\widehat{\cal M}(\omega)$
under $T$-duality. Namely, as is well known, one can start from
background configurations where the NS-NS field $B_{ij}=0$ and
generate new backgrounds with $B_{ij}\ne 0$ through a 
$T$-duality rotation. Correspondingly, we can construct $\widehat{\cal
 M}^{(B=0)}(\omega)$ explicitly and generate $\widehat{\cal
 M}^{(B)}(\omega)$ through a $T$-duality rotation. Alternatively, we
 can also construct the $\widehat{\cal M}^{(B)}(\omega)$
 directly and demonstrate that the two methods indeed lead to identical
 results.  

Our starting point is the $D$-dimensional tree level string effective action
\bea
\label{tree}
{\hat S}=\int d^Dx\,\sqrt {-\widehat{G}}\left(R_{\widehat{G}} +({\hat \partial}
{\hat \phi})^2 -{1\over {12}}{\widehat H}^2 \right) \eea
where $\widehat{G}_{{\hat \mu}{\hat \nu}}$ is the metric in
$D$-dimensions in the string frame, $\widehat{G}$ its determinant,
$R_{\widehat G}$ the corresponding scalar curvature,
$\hat \phi$ is the dilaton and ${\widehat H}=d{\widehat B}$. When compactified
on $T^d$, the reduced action takes the following form \cite{jmj}:
 \bea
\label{twod}
S=\int dx^0dx^1\,{\sqrt {-g}}e^{-\bar {\phi}}\left(R+(\partial {\bar \phi})^2+
{1\over 8}{\rm Tr}(\partial _{\alpha}M^{-1}\partial ^{\alpha}M)
\right) \eea
where $g_{\alpha \beta}$ with $\alpha ,\beta =0,1$ 
is the two dimensional space-time metric, and $R$ is the corresponding
scalar curvature.
The shifted dilaton is  ${\bar \phi}=
\phi -{1\over 2} {\rm log}~{\rm det}~G $,  with $G_{ij}$  the
metric corresponding to the transverse coordinates $x^i, i=2,3.., D-1$. $M$ is
a $2d \times 2d$ symmetric matrix belonging to $O(d,d)$,  
\be
\label{mmatrix}
M = \pmatrix {G^{-1} & -G^{-1} B\cr
BG^{-1} & G - BG^{-1} B\cr} \ee
where $B$ represents the moduli coming from the reduction of the
$B$-field  in $D$ space-time
dimensions. The symmetric nature of $M$ is evident since $G$ is
symmetric and $B$  is anti-symmetric.
In general, there will be additional terms in (\ref{twod})
corresponding to $d$ Abelian gauge fields coming from  the original
metric and another set of $d$ gauge fields coming from the
anti-symmetric tensor $B_{\hat{\mu}\hat{\nu}}$ as a
result of dimensional reduction \cite{jmj} ($d=D-2$). 
Furthermore, there  would also have been the field
strength of the two dimensional tensor field $B_{\alpha\beta}$. Since we are
effectively in two space-time dimensions, we have dropped the gauge field terms
and, in the same spirit, have not kept the field strength of $B_{\alpha\beta}$,
which can be removed if it depends only on the coordinates $x^0$ and $x^1$.

The action (\ref{twod}) is
invariant  under the global $O(d,d)$ transformations,
\be g_{\alpha\beta} \rightarrow g_{\alpha\beta}, ~~ 
 {\bar \phi} \rightarrow {\bar \phi},\qquad M \rightarrow \Omega ^TM\Omega \ee
where $\Omega \in O(d,d)$ is the transformation matrix, which leaves
the $O(d,d)$ metric, $\eta={\scriptstyle \pmatrix {0 & {\bf 1}\cr {\bf
1} & 0\cr}}$, 
invariant ($\bf 1$ is the $d$-dimensional unit matrix). It is
convenient to introduce the lower triangular matrix
\bea
\label{v}
V=\pmatrix {E^{-1} & 0 \cr BE^{-1} & E^T \cr } \eea
where $E$ is the vielbein with $(EE^T)_{ij}=G_{ij}$. It is easy to
check that  $V\in O(d,d)$ and that 
$M=VV^T$ so that under
$O(d,d)$ transformations,    
$V \rightarrow \Omega ^T V $. In fact, $V$ belongs to the coset
${O(d,d)\over O(d)\times O(d)}$.

We may decompose the current as
\bea
\label{current}
V^{-1}\partial _{\alpha} V=P_{\alpha}+Q_{\alpha} \eea
where $Q_{\alpha} \in O(d)\times O(d)$,
the Lie algebra for the maximally compact subgroup and $P_{\alpha}$
belongs to the complement. It follows from the property of the coset
under study that $P_{\alpha}^{T} = P_{\alpha}$ and $Q_{\alpha}^{T} = -
Q_{\alpha}$ and it is straightforward to check that the
last term  of (\ref{twod}) can be expressed as
\be
\label{eqp}
 {\rm Tr}~(\partial _{\alpha}M\partial ^{\alpha}M^{-1})= -4{\rm Tr}~
g^{\alpha\beta}P_{\alpha}P_{\beta} \ee
Let us note
that, under a global $O(d,d)$ rotation $V\rightarrow \Omega^{T}V$, the
current in (\ref{current}) is invariant, while
under a local $O(d)\times O(d)$ transformation $V
\rightarrow Vh(x)$, $P_{\alpha}$ and $Q_{\alpha}$
transform as
\be P_{\alpha} \rightarrow h^{-1} (x)P_{\alpha}h(x),~~~
Q_{\alpha}  \rightarrow
h^{-1}(x)Q_{\alpha}h(x)+h^{-1}(x)\partial_{\alpha}h(x) \ee
which leaves (\ref{eqp}) invariant.

A few comments are in order at this stage. We are dealing with a
2-dimensional $\sigma$-model \cite{ib,jhs,sen,mizo} coupled to gravity
and,  in addition, 
the action (\ref{twod}) contains a shifted dilaton field, $\bar \phi$.
We may rescale the metric $g_{\alpha\beta}\rightarrow e^{\bar
\phi}g_{\alpha\beta}$, which will eliminate the kinetic energy term of $\bar
\phi$. Recall that in a curved space, the spectral parameter becomes
space-time dependent in the zero curvature formulation of the sigma
model, unlike in flat space, where the spectral parameter is a
constant. Let ${\widehat V}(x,t)$,
where $t$ is the space-time dependent spectral parameter, be the one
parameter family of matrices such that $\widehat{V}(x,t=0) =
V(x)$ and 
\bea
\label{spectral}
{\widehat V}^{-1}(x,t)\partial _{\alpha}{\widehat
V}(x,t) = Q_{\alpha}+{{1+t^2}\over{1-t^2}}P_{\alpha} +{{2t}\over
{1-t^2}}\epsilon _{\alpha\beta}P^{\beta} \eea
Then, it is easy to check that the integrability condition
\bea
\label{int}
\partial_{\alpha}({\widehat V}^{-1}\partial_{\beta}{\widehat V})-\partial
_{\beta} ({\widehat V}^{-1}\partial _{\alpha}{\widehat V})+[{\widehat V}^{-1}
\partial _{\alpha}\widehat{V} , {\widehat V}^{-1}\partial
_{\beta}{\widehat V}]=0 \eea
leads to the dynamical equations of the theory, provided the spectral
parameter is related to the shifted dilaton through the equation
\be
\label{teq}
\partial _{\alpha}t=-{1\over 2}\epsilon _{\alpha\beta}\partial
^{\beta}\left(e^{-\bar \phi}(t+{1\over t})\right) \ee
The solution to this equation has the form
\be
t(x) = {\sqrt{\omega + \rho_{+}(x^{+})} - \sqrt{\omega -
\rho_{-}(x^{-})}\over \sqrt{\omega + \rho_{+}(x^{+})} + \sqrt{\omega -
\rho_{-}(x^{-})}} \label{double}
 \ee
where we have defined $\rho = e^{-\bar{\phi}} = \rho_{+}(x^{+}) +
\rho_{-}(x^{-})$ for later use and
$\omega$ is  the constant of integration (which can be thought of as a
global, constant spectral parameter). The light cone coordinates are
defined to be 
$x^{+}={{1\over {\sqrt 2}}(x^0+x^1)}$ and $x^{-}={{1\over {\sqrt
2}}(x^0-x^1)}$. 

The next important step is to solve for the monodromy matrix,
$\widehat{\cal M}(\omega)$. The linear system (\ref{spectral}) is
known to be invariant under a generalization of the symmetric space
automorphism, $\tau ^{\infty} $, defined through 
\be \tau ^{\infty}{\widehat V}(x,t)=\left({\widehat
V}^{-1}\right)^{T}(x,{1\over t}) \ee
This relation has the following action on  the elements of the Lie
algebra,  $Q_{\alpha}\rightarrow Q_{\alpha}$,
$P_{\alpha}\rightarrow - P_{\alpha}$ and $t\rightarrow
{1\over t}$. The monodromy matrix, defined as,
\be
\label{mono}
\widehat{\cal M}(\omega) = {\widehat V}(x, t)\tau^{\infty}{\widehat
V}^{-1} (x,t) = {\widehat V}(x,t){\widehat V}^T (x, {1\over t}) 
\ee
is space-time independent and plays a cardinal role in the study of
integrable models. The
reconstruction of solutions and properties of integrability are all deeply
encoded in $\widehat{\cal M}(\omega)$. For example, if we were to use
a  solution of the $\bar{\phi}$ equation as input, then we can
determine $t$ from its relation with the dilaton and eventually obtain
$\widehat{V}(x,t)$.

Now we want to present a crucial observation which enables us to
derive  the
transformation property of $\widehat{\cal M}(\omega)$ under
$T$-duality. Note that the one parameter family of matrices 
$\widehat{V}(x,t)\in {O(d,d)\over O(d)\times O(d)}$ much like $V(x) =
\widehat{V}(x,t=0)$. Therefore, under a global $O(d,d)$ rotation,
$\widehat{V}(x,t)\rightarrow \Omega^{T}\widehat{V}(x,t)$ and the
monodromy matrix transforms as
\be
\widehat{\cal M}(w) \rightarrow \Omega ^T \widehat{\cal M}(w)\Omega 
\ee
Furthermore, although under a local $O(d)\times O(d)$ transformation
$\widehat{V}(x,t)\rightarrow \widehat{V}(x,t) h(x)$, the only local
transformations, which will preserve the global nature of the
monodromy matrix are the ones which do not depend on the spectral
parameter explicitly. This is very much like what we already know for
the case $t=0$, namely,  
although $V\rightarrow \Omega ^T Vh(x)$ under combined $O(d,d)$ and
$O(d)\times O(d)$ transformations,  $M=VV^{T}$  undergoes
only a  global $O(d,d)$ rotation.
In other words, the monodromy matrix, $\widehat{\cal M}(\omega)$, is
only affected by the global $O(d,d)$ rotation, as is the case with the
$M$-matrix. Thus, we are able to synthesize the integrability properties
of the two dimensional string effective action with its $T$-duality
properties. Indeed, this relation can be explicitly checked in
specific examples (as we will show shortly) and can be used as a
powerful tool  in determining solutions. Let us derive the
transformation  properties of $\widehat{\cal M}$ under
infinitesimal $O(d,d)$ transformations. We may express $\widehat{\cal M}$ as
\bea
\label{block}
\widehat{\cal M}=\pmatrix {\widehat{\cal M}_{11} & \widehat{\cal
M}_{12} \cr  
\widehat{\cal M}_{21} & \widehat{\cal M}_{22} \cr } \eea
where each element is a $d\times d$ matrix. The matrix $\Omega$
assumes the following form
\bea
\label{omega}
\Omega = \pmatrix {1+X & Y \cr Z & 1+W \cr } \eea
where the infinitesimal parameters of the transformation are required to
satisfy, $Y^T=-Y, Z^T=-Z$ and $W=-X^T$. Under (\ref{omega}), the elements of
(\ref{block}) transform as
\begin{eqnarray}
\label{mtr}
\delta \widehat{\cal M}_{11} & = &\widehat{\cal
M}_{11}X+X^T\widehat{\cal M}_{11}-Z\widehat{\cal M}_{12}+
\widehat{\cal M}_{12}Z\nonumber\\
\delta \widehat{\cal M}_{12} & = & \widehat{\cal
M}_{11}Y+X^T\widehat{\cal M}_{12}-Z \widehat{\cal
M}_{22}-\widehat{\cal M}_{12}X^T\nonumber\\ 
\delta \widehat{\cal M}_{21} & = & -Y\widehat{\cal
M}_{11}-X\widehat{\cal M}_{21}+{\cal M}_{21}X+\widehat{\cal
M}_{22}Z\nonumber\\ 
 \delta \widehat{\cal M}_{22} & = & \widehat{\cal M}_{21}Y- Y{\cal
M}_{12} -X\widehat{\cal M}_{22}-\widehat{\cal M}_{22}X^T 
\end{eqnarray}
 
Since, we can obtain $\widehat{\cal M}(\omega)$ for the case when
$B\ne 0$ starting from a vanishing $B$ configuration, let us derive
$\widehat{\cal M}(\omega)$ for the case when $B=0$, $E$ is diagonal so
that $G_{ij}$ 
is also diagonal. It is easy to check that $Q_{\alpha}=0$ for this
choice of backgrounds and the relevant equation for determining
$\widehat{V}^{(B=0)}$  are given by
\be
\label{diag}
({\widehat V}^{(B=0)})^{-1}\partial_{+} {\widehat
V}^{(B=0)} = {{1-t}\over {1+t}}\,P_{+} ~~{\rm and}~~
({\widehat V}^{(B=0)})^{-1}\partial_{-} {\widehat
V}^{(B=0)} = {{1+t}\over {1-t}}\,P_{-} \ee
and $t$ satisfies (\ref{double}). 
The factorizability property of $\widehat{\cal M}(\omega)$, namely  
(\ref{mono}), together with the assumption that it has only isolated
single poles determines that the monodromy matrix (as
well as the $\widehat{V}^{(B=0)}(x,t)$ matrix)  will
have $2d$ poles in the spectral parameter. We can check this in the
following manner. Let us assume that $G_{ij}$ has the diagonal structure
\begin{equation}
\label{gmatrix}
G= {\rm diag}\,(e^{\lambda +\psi_1}, e^{\lambda +\psi_2},\cdots ,e^{\lambda
+\psi_d} )
\end{equation}
with $\sum \psi_i=0$, so that $\lambda =\frac{1}{d}\log{\det{G}}$ as
adopted in \cite{bv}.  
The $E$-matrix is also diagonal and has the form
\begin{equation}
E = {\rm diag}\,(e^{{1\over 2}(\lambda + \psi_{1})}, e^{{1\over
2}(\lambda + \psi_{2})},\cdots , e^{{1\over 2}(\lambda +
\psi_{d})})\label{ematrix} 
\end{equation}
Correspondingly, the $\widehat{V}^{(B=0)}$ matrix is diagonal as well
and has  the form
\begin{equation}
\widehat{V}^{(B=0)}(x,t) = \left(\begin{array}{cc}
\overline{V}^{(B=0)}(x,t) & 0\\
0 & (\overline{V}^{(B=0)})^{-1}(x,t)
\end{array}\right)
\end{equation}

Let us assume the following form for the diagonal matrix
$\overline{V}^{(B=0)}(x,t) = (\overline{V}_{1},\overline{V}_{2},\cdots , \overline{V}_{d})$ with
\begin{equation}
\overline{V}_{i} = {t_{d+i}\over t_{i}} \,{t-t_{i}\over t-t_{d+i}}\,E_{i}^{-1}
\end{equation}
which explicitly brings out the $2d$ pole structures in
$\widehat{V}^{(B=0)}(x,t)$.  The spectral parameters satisfy
\begin{equation}
\partial_{\pm} t = {1\mp t\over 1\pm t}\,\partial_{\pm} \ln
\rho,\qquad \partial_{\pm}t_{i} = {1 \mp t_{i}\over 1\pm
t_{i}}\,\partial_{\pm} \ln \rho
\end{equation}
Using this, it is easy to verify that ($t_{d+i}$ and $t_{i}$ have
opposite signatures, following from the double valued nature of the
solutions in Eq. (\ref{double}))
\begin{equation}
\overline{V}_{i}^{-1}\partial_{\pm} \overline{V}_{i} = \partial_{\pm} \ln E_{i}^{-1} \mp
{t\over 1 \pm t}\,\partial_{\pm}\ln \left(-{t_{i}\over t_{d+i}}\right) = {1\mp
t\over 1\pm t}\, \partial_{\pm}\ln E_{i}^{-1}
\end{equation}
provided we identify $- {t_{i}\over t_{d+i}} = E_{i}^{-2}$. In such a
case,  we can write
\begin{equation}
(\widehat{V}^{(B=0)})^{-1}(x,t)\partial_{\pm}
\widehat{V}^{(B=0)}(x,t) = {1\mp t\over 1\pm
t}\left(\begin{array}{cc}
- E^{-1}\partial_{\pm}E & 0\\
0 & E^{-1}\partial_{\pm} E
\end{array}\right) = {1\mp t\over 1\pm t}\,P_{\pm}
\end{equation}

Thus, we see that, for the case at hand, we have to introduce $2d$ poles, a
pair for every diagonal element $E_{i}$. Furthermore, the diagonal elements of
$\overline{V}^{(B=0)}(x,t)$, in this case, are determined to be
\begin{equation}
\overline{V}_{i}(x,t) = {t_{d+i}\over t_{i}}\,{t-t_{i}\over t-t_{d+i}}\,E_{i}^{-1}
= \sqrt{-{t_{d+i}\over t_{i}}}\, {t-t_{i}\over t-t_{d+i}}\label{V}
\end{equation}
It is easy to check that the global (constant) spectral parameters satisfy
\begin{equation}
{\omega - \omega_{i}\over \omega - \omega_{d+i}} = {t_{d+i}\over
t_{i}}\,{t-t_{i}\over t-t_{d+i}}\,{{1\over t} - t_{i}\over {1\over t}
- t_{d+i}}
\end{equation}
Using this relation as well as the definition of the monodromy matrix, we note
that we can write
\begin{equation}
\widehat{\cal M}^{(B=0)} (\omega) = \widehat{V}^{(B=0)}(x,t)
(\widehat{V}^{(B=0)})^{T}(x,{1\over t}) = 
\left(\begin{array}{cc} 
{\cal M}(\omega) & 0\\
0 & {\cal M}^{-1}(\omega)
\end{array}\right)
\end{equation}
where ${\cal M}(\omega)$ is diagonal with
\begin{equation}
{\cal M}_{i}(\omega) = \overline{V}_{i}(x,t) \overline{V}_{i}(x,{1\over t})
= - {\omega - \omega_{i}\over \omega - \omega_{d+i}}
\end{equation}
The double valued relation between the global and the local spectral
parameters allows us to choose $\omega_{d+i} =
-\omega_{i}$, which leads to
\begin{equation}
{\cal M}_{i}(\omega) = {\omega_{i} - \omega\over \omega_{i} +
\omega}\label{Mono}
\end{equation}
This determines the monodromy matrix for a pure $G$ background and
starting from this, we
can determine the monodromy matrix for a background with a
non-vanishing $G$  and $B$ generated through an  $O(d,d)$
transformation.

We shall adopt the Nappi-Witten model to illustrate these points
explicitly. The Nappi-Witten model is described by the following 
${SU(2)\over U(1)}\times {SL(2,R)\over SO(1,1)}$ gauged WZW model
\cite{nw}. The worldsheet action, in terms of holomorphic and
anti-holomorphic coordinates, is:
\begin{equation}
S  = \int d^{2}z( -\partial \psi \overline{\partial }
\psi +\partial s\overline{\partial }s+\partial \rho \overline{\partial }\rho
{G}_{\rho \rho}
 +\partial \lambda \overline{\partial }\lambda{G}_{\lambda \lambda} 
+(\partial \rho \overline{\partial }\lambda -\partial \lambda 
\overline{\partial }\rho ) B )
\end{equation}
where (we will identify $x^{0}=\tau$ here)
\begin{equation}
G_{\rho \rho } = \frac{4\cos ^{2}\tau\cos ^{2}x}{(1-\cos 2\tau\cos 2x)},\qquad
G_{\lambda \lambda } = \frac{4\sin^{2}\tau\sin ^{2}x}{(1-\cos 2\tau\cos
2x)},\qquad
B = \frac{(\cos 2\tau-\cos 2x)}{(1-\cos 2\tau\cos
2x)}\label{background} 
\end{equation}
and $\bar{\phi} = - \ln \left(\sin 2x \sin 2\tau\right)$. This model
describes  a $4$-dimensional cosmological scenario with
graviton, dilaton and anti-symmetric tensor fields and
possesses two isometries.

The backgrounds, in Eq. (\ref{background}), can be generated from the
ones where $B=0$ and the  metric given by:
\begin{eqnarray}
ds^2=-d\tau^2+dx^2+\tan^{-2}\tau\,dy^2+\tan^2x\,dz^2
\end{eqnarray}
through an O(2,2) transformation (see \cite{gmv} for a more general
discussion):
\begin{equation}
\Omega =\frac{1}{\sqrt{2}}\left( 
\begin{array}{cccc}
1 & 0 & 0 & 1 \\ 
0 & 1 & -1 & 0 \\ 
0 & 1 & 1 & 0 \\ 
-1 & 0 & 0 & 1
\end{array}
\right)  \label{Omega}
\end{equation}
We can construct $\cal{M}(\omega)$ corresponding to these two vacua following 
the procedure described earlier (see Eqs. (\ref{V},\ref{Mono}))
\begin{equation}
\widehat{V}^{(B=0)}(x,t)= {\rm diag}({\scriptstyle 
\sqrt{-\frac{t_{3}}{t_{1}}}\frac{t-t_{1}}{t-t_{3}},  
\sqrt{-\frac{t_{4}}{t_{2}}}\frac{t-t_{2}}{t-t_{4}}, 
\sqrt{-\frac{t_{1}}{t_{3}}}\frac{t-t_{3}}{t-t_{1}}, 
\sqrt{-\frac{t_{2}}{t_{4}}}\frac{t-t_{4}}{t-t_{2}}})\label{Vee}
\end{equation}
Thus, the monodromy matrix for $B=0$ follows to have the form 
\begin{equation}
\widehat{\cal M}^{(B=0)}(\omega)=\left( 
\begin{array}{cccc}
\frac{\omega _{1}-\omega }{\omega _{1}+\omega } & 0 & 0 & 0 \\ 
0 & \frac{\omega _{2}-\omega }{\omega _{2}+\omega } & 0 & 0 \\ 
0 & 0 & \frac{\omega _{1}+\omega }{\omega _{1}-\omega } & 0 \\ 
0 & 0 & 0 & \frac{\omega _{2}+\omega }{\omega _{2}-\omega }
\end{array}
\right) = \left(\begin{array}{cc}
{\cal M}(\omega) & 0\\
0 & {\cal M}^{-1}(\omega)
\end{array}\right) \label{mono4}
\end{equation}

We can also compute $\widehat{V}^{(B)}(x,t)$ directly from the
Nappi-Witten  action which takes the form:
\begin{equation}
\widehat{V}^{(B)}(x,t)=\frac{1}{\sqrt{2}}\left( 
\begin{array}{cccc}
\gamma \hat{V}_{1}+\delta \hat{V}_{4} & 0 & 0 & \delta
\hat{V}_{1}-\gamma \hat{V}_{4} \\ 
0 & \gamma \hat{V}_{2}+\delta \hat{V}_{3} & -\delta \hat{V}_{2}+\gamma
\hat{V}_{3} & 0 \\ 
0 & -\gamma \hat{V}_{2}+\delta \hat{V}_{3} & \delta \hat{V}_{2}+\gamma
\hat{V}_{3} & 0 \\ 
\gamma \hat{V}_{1}-\delta \hat{V}_{4} & 0 & 0 & \delta
\hat{V}_{1}+\gamma \hat{V}_{4}
\end{array}
\right)  \label{Vprima}
\end{equation}
where $\gamma=\sqrt{\frac{1-B}{2}}$, $\delta=\sqrt{\frac{1+B}{2}}$,
and $\hat{V}_{i}$ are the diagonal elements of
$\widehat{V}^{(B=0)}(x,t)$ in (\ref{Vee}). 
This leads to
\begin{equation}
\widehat{\cal M}^{(B)}(\omega)=\frac{1}{2}\left(
\begin{array}{cccc}
{\cal M}_{1}+{\cal M}^{-1}_{2} & 0 & 0 & {\cal M}_{1}-{\cal M}^{-1}_{2} \\
0  & {\cal M}_{2}+{\cal M}^{-1}_{1} & -{\cal M}_{2}+{\cal M}^{-1}_{1} & 0 \\ 
0 & {\cal M}_{2}+{\cal M}^{-1}_{1} & {\cal M}_{2}+{\cal M}^{-1}_{1} & 0 \\
{\cal M}_{1}-{\cal M}^{-1}_{2} & 0 & 0 & {\cal M}_{1}+{\cal M}^{-1}_{2} \\
\end{array}
\right)
\end{equation}
where ${\cal M}_{1}=\frac{\omega _{1}-\omega }{\omega _{1}+\omega}$ and 
${\cal M}_{2}=\frac{\omega _{2}-\omega }{\omega _{2}+\omega}$ are the
two diagonal elements of ${\cal M}(\omega)$ in (\ref{mono4}).
It can now be explicitly checked that $\widehat{\cal M}^{(B)}(\omega)$ and 
$\widehat{\cal M}^{(B=0)}(\omega)$ 
are related by the $\Omega$ matrix as was argued 
earlier. Furthermore, it is also straightforward to check that
$\widehat{V}^{(B)}(x,t) = \Omega^{T}\widehat{V}^{(B=0)}(x,t)h(x)$ where
\begin{equation}
h(x) = \left(\begin{array}{rrrr}
\gamma & 0 & 0 & \delta\\
0 & \gamma & -\delta & 0\\
0 & \delta & \gamma & 0\\
-\delta & 0 & 0 & \gamma
\end{array}\right)
\end{equation}
is independent of the spectral parameter as was pointed out earlier. 

To summarize our results, it is important to note that we have
revealed a deep connection between duality symmetry of two dimensional
string 
effective action and its integrability properties through the construction
of the monodromy matrix for the general case. Furthermore, we have
shown that $\widehat{\cal M}(\omega)$ transforms nontrivially under the
$O(d,d)$ transformations. Our results are likely to have applications
in a variety of problems in string theory which are described by an
effective two dimensional action (such as collisions of plane waves in
the study of blackholes, cosmological scenarios etc). We have
presented an  application of our
results in the study of an exactly solvable WZW model \cite{nw}, which
unravels some of the salient features of our investigation. As is evident,
a wide class of problems can be studied from this perspective.

\vskip .7cm
\noindent {\bf Acknowledgment:} One of us (JM) acknowledges the warm
hospitality of Prof. Y. Kitazawa and KEK. This work is supported in part
by US DOE Grant No. DE-FG 02-91ER40685.

%\centerline{{\bf References}}

\end{document}